# The anisotropy induced by defects of "random local field" type in $O(n)$ models and suppression of the Imry-Ma inhomogeneous state

*A.A. Berzin[1], A.I. Morosov[2*], and A.S. Sigov[1]*

[1] Moscow Technological University (MIREA), 78 Vernadsky pr., Moscow, 119454, Russian Federation;
[2] Moscow Institute of Physics and Technology (State University), 9 Institutskiy per., Dolgoprudny, Moscow Region, 141700, Russian Federation.

## Abstract

We demonstrate that in the system with anisotropic distribution of the defect-induced random local field directions in the $n$-dimensional space of vector order parameter with the $O(n)$ symmetry, the defect-induced effective anisotropy arises for the space dimensionality $2<d<4$. If the anisotropy constant exceeds the threshold value, then an inhomogeneous state predicted by Imry and Ma becomes energetically unfavorable, and the system goes back to the state with the long-range order.

[*] E-mail: mor-alexandr@yandex.ru

## 1. Introduction

After the publication in 1975 the classical paper by Imry and Ma [1], the viewpoint was firmly established in the literature that in space dimensions $d<4$ the introduction of an arbitrarily small concentration of defects of the "random local field" type in a system with continuous symmetry of the $n$-component vector order parameter ($O(n)$ model) leads to the long-range order collapse and to the occurrence of an inhomogeneous state, which in what follows will be designated as the Imry-Ma state.

However, as shown in paper [2], the arguments [1] on the impossibility of the existence of the long range order in space $d<4$, in reality are true when $d\leq 2$, that is, when the long-range order does not exist at finite temperature even in a perfect system (without defects). In space dimensions $2<d<4$, and in particular, in the three-dimensional space, an inhomogeneous state can, generally speaking, coexist with the long-range order.

In the present paper we demonstrate that the anisotropic distribution of the defect-induced random local field directions in the $n$-dimensional space of the order parameter in most cases leads to the impurity-induced effective anisotropy in the system. It was shown previously [3] that if the anisotropy constant exceeds a certain threshold value, then the Imry-Ma inhomogeneous state becomes energetically unfavorable, and the system goes back to the state with the long-range order. The comparison of the specified effective anisotropy constant to its threshold value indicates that the Imry-Ma state does not occur in most cases of anisotropic distribution of the random local field directions.

## 2. Energy of a system of classical spins

The exchange interaction energy of $n$-component localized spins $S_i$ comprising the $d$-dimensional lattice has the form

$$W_{ex} = -\sum_{i,j>i} J_{ij} S_i S_j, \tag{1}$$

where $J_{ij}$ is the exchange interaction constant for *i*-th and *j*-th spins, and the summation is performed over the whole spin lattice.

The energy of interaction between the spins and defect-induced random local fields is

$$W_{imp} = -\sum_l \boldsymbol{S}_l \boldsymbol{h}_l, \qquad (2)$$

where the summation is performed over defects randomly located in the lattice sites, and the density of random local fields $\boldsymbol{h}$ distribution in the spin space (order parameter space) possesses the property $\rho(\boldsymbol{h}) = \rho(-\boldsymbol{h})$, which provides the lack of mean field in an infinite system.

Switching to the continuous distribution of the order parameter $\boldsymbol{\eta}$, we introduce the energy of inhomogeneous exchange in the form [4]

$$\widetilde{W}_{ex} = -\frac{1}{2}\int d^d\boldsymbol{r}\, D\, \frac{\partial \eta^\perp}{\partial x_i}\frac{\partial \eta^\perp}{\partial x_i}, \qquad (3)$$

where $\boldsymbol{\eta} \sim \boldsymbol{S}_l/b^d$, $b$ is the interstitial distance, $D \sim Jb^{2+d}$, $J$ is the exchange integral reflecting the interaction of the nearest neighbors, and $\boldsymbol{\eta}^\perp(\boldsymbol{r})$ is the order parameter component orthogonal to the direction of the order parameter $\boldsymbol{\eta}_0$ in the system without defects.

The energy of interaction between the random field $\boldsymbol{h}(\boldsymbol{r})$ and the order parameter $\boldsymbol{\eta}(\boldsymbol{r})$ has the form

$$W_{def} = -\int d^d\boldsymbol{r}\, \boldsymbol{h}(\boldsymbol{r})\boldsymbol{\eta}(\boldsymbol{r}), \qquad (4)$$

where

$$\boldsymbol{h}(\boldsymbol{r}) = b^d \sum_l \boldsymbol{h}_l\, \delta(\boldsymbol{r} - \boldsymbol{r}_l). \qquad (5)$$

### 3. Effective anisotropy

For a qualitative explanation of the mechanism of the effective anisotropy occurrence let us consider the influence of defect-induced local field $\boldsymbol{h}_l$ upon uniform distribution of the order parameter. Thus, for simplicity, we neglect the longitudinal susceptibility of the system at low temperatures, much smaller than the temperature of magnetic ordering.

The component of the random field $h_l^\perp$ perpendicular to the $\eta_0$ direction in the system free of defects leads to a local deviation of the order parameter and to the appearance of the orthogonal component $\eta^\perp(r)$. The result is a negative additive to the energy of the ground state proportional to $(h_l^\perp)^2$. It is maximum in modulus when the $\eta_0$ direction is perpendicular to the defect-induced local field.

In the particular case of anisotropic distribution of the random fields directions, when all $h_l$ are collinear, the order parameter is energetically favorable to be oriented perpendicularly to this direction. Thus, in the case of X-Y model (*n*=2) there arises anisotropy of the "easy axis" type, and in the case of the Heisenberg model (*n*=3) one obtains anisotropy of the "easy plane" type.

In the case of coplanar distribution of the random field directions in the space of the order parameter in the Heisenberg model, there arises an easy axis perpendicular to the indicated plane.

For a more general anisotropic distribution of the random field directions, the order parameter is advantageous to be oriented perpendicularly to the preferential direction of the random fields.

Let us find the expression for the anisotropy energy in the case of arbitrary distribution of the directions of the defect-induced random fields. We represent the order parameter in the linear approximation for $h$ in the form

$$\eta(r) = \eta_0 + \eta^\perp(r), \tag{6}$$

where $|\eta_0| \gg |\eta^\perp(r)|$.

The term proportional to $\eta_0$ in the expression for $W_{def}$ vanishes because of the function $\rho(h)$ parity.

The quantity $h^\perp(r)$ may be represented by a sum

$$h^\perp(r) = b^d \sum_l [h_l - n(nh_l)] \delta(r - r_l), \tag{7}$$

where $\mathbf{n} = \boldsymbol{\eta}_0/|\boldsymbol{\eta}_0|$.

The Fourier component of the function $\boldsymbol{\eta}^\perp(\mathbf{k})$ is related to the Fourier component of the random field $\mathbf{h}^\perp(\mathbf{k})$:

$$\boldsymbol{\eta}^\perp(\mathbf{k}) = \chi^\perp(\mathbf{k})\mathbf{h}^\perp(\mathbf{k}), \qquad (8)$$

where

$$\chi^\perp(\mathbf{k}) = (Dk^2)^{-1}, \qquad (9)$$

whereas for the quantity $\mathbf{h}^\perp(\mathbf{k})$ one can write down

$$\mathbf{h}^\perp(\mathbf{k}) = \frac{1}{V}\int d^d\mathbf{r}\, \mathbf{h}^\perp(\mathbf{r})\exp(-i\mathbf{k}\mathbf{r}) = \frac{1}{N}\sum_l [\mathbf{h}_l - \mathbf{n}(\mathbf{n}\mathbf{h}_l)]\exp(-i\mathbf{k}\mathbf{r}_l), \qquad (10)$$

where $N$ is the number of elementary cells. Then one obtains from Eqs. (8) and (10)

$$\boldsymbol{\eta}^\perp(\mathbf{r}) = \frac{1}{N}\sum_\mathbf{k}\chi^\perp(\mathbf{k})\sum_l[\mathbf{h}_l - \mathbf{n}(\mathbf{n}\mathbf{h}_l)]\exp[i\mathbf{k}(\mathbf{r}-\mathbf{r}_l)], \qquad (11)$$

and the quantity $W_{def}$ takes the form

$$W_{def} = -\frac{1}{N^2}\int d^d\mathbf{r}\sum_{\mathbf{k},\mathbf{k}'}\chi^\perp(\mathbf{k})\sum_{l,m}[\mathbf{h}_l - \mathbf{n}(\mathbf{n}\mathbf{h}_l)][\mathbf{h}_m - \mathbf{n}(\mathbf{n}\mathbf{h}_m)] \times$$

$$\times \exp[i\mathbf{k}(\mathbf{r}-\mathbf{r}_l) + i\mathbf{k}'(\mathbf{r}-\mathbf{r}_m)], \qquad (12)$$

The integration produces the factor $V\delta_{\mathbf{k},-\mathbf{k}'}$. Ultimately we have

$$W_{def} = -\frac{V}{N^2}\sum_\mathbf{k}\chi^\perp(\mathbf{k})\sum_{l,m}[\mathbf{h}_l - \mathbf{n}(\mathbf{n}\mathbf{h}_l)][\mathbf{h}_m - \mathbf{n}(\mathbf{n}\mathbf{h}_m)] \times$$

$$\times \exp[i\mathbf{k}(\mathbf{r}_m - \mathbf{r}_l)]. \qquad (13)$$

Due to random distribution of defects in the coordinate space, the contribution different from zero results from the summands with $l=m$. Therefore Eq. (13) yields

$$W_{def} = -\frac{V}{N^2}\sum_\mathbf{k}\chi^\perp(\mathbf{k})\sum_l[\mathbf{h}_l - \mathbf{n}(\mathbf{n}\mathbf{h}_l)]^2 =$$

$$= -xb^d \sum_{\mathbf{k}} \chi^{\perp}(\mathbf{k}) \langle [\mathbf{h}_l - \mathbf{n}(\mathbf{n}\mathbf{h}_l)]^2 \rangle. \tag{14}$$

Here $x$ is the dimensionless concentration of defects (the number of defects per a unit cell), and the brackets $\langle \rangle$ denote averaging over all defect local fields. Going from summation over $\mathbf{k}$ to integration over the Brillouin zone and introducing the notation

$$\tilde{\chi}^{\perp} = b^d \int \frac{d^d \mathbf{k}}{(2\pi)^d} \chi^{\perp}(\mathbf{k}), \tag{15}$$

we get the volume density of the energy of interaction between the order parameter and defect-induced random local fields

$$w_{def} = -x\tilde{\chi}^{\perp}\left[\langle \mathbf{h}_l^2 \rangle - \langle (\mathbf{n}\mathbf{h}_l)^2 \rangle\right]. \tag{16}$$

In space dimension $2<d<4$ the quantity $\tilde{\chi}^{\perp}$ has no singularities at $\mathbf{k} = 0$.

It can be easily seen that in the case of anisotropic distribution of the directions of random fields the second term in square brackets on the right side of Eq. (16) induces the anisotropy in the order parameter space. In particular, for the collinear orientation of random fields, the volume density of anisotropy energy takes the form

$$w_{an} = x\tilde{\chi}^{\perp}\langle \mathbf{h}_l^2 \rangle \cos^2\varphi \equiv \frac{1}{2} K_{eff} S^2 b^{-d} \cos^2\varphi, \tag{17}$$

where $\varphi$ is the angle between the order parameter vector and the axis of "hard magnetization" (defect-induced random local fields are collinear to this axis), and $S$ is the magnitude of the spin vector.

In the case of coplanar and isotropic in the selected plane distribution of random fields in the Heisenberg model the volume density of anisotropy energy is

$$w_{an} = -\frac{1}{2}x\tilde{\chi}^{\perp}\langle \mathbf{h}_l^2 \rangle \cos^2\varphi \equiv -\frac{1}{2} K_{eff} S^2 b^{-d} \cos^2\varphi, \tag{18}$$

$\varphi$ being the angle between the order parameter vector and normal to the plane containing random field vectors.

Evaluation of the effective anisotropy constant $K_{eff}$ in the order of magnitude gives the value

$$K_{eff} \sim \frac{x\langle h_l^2\rangle}{JS^2}. \tag{19}$$

In the general case of anisotropic distribution of the directions of random fields, it is convenient to describe the given anisotropy through the difference $\Delta$ between the maximum and minimum value of the expression $\langle(\mathbf{n}, \mathbf{h}_l)^2\rangle$ as a function of the vector $\mathbf{n}$ direction:

$$w_{an} = x\tilde{\chi}^\perp \Delta = \frac{1}{2} K_{eff} S^2 b^{-d}. \tag{20}$$

### 4. Suppression of Imry-Ma inhomogeneous state

Let us compare the obtained effective anisotropy constant with its critical value. If the effective anisotropy constant exceeds the critical value then the Imry-Ma inhomogeneous state shall be suppressed [3]. Indeed, to follow the fluctuations of the random field, the order parameter has to deviate from the most favorable (from the point of view of the anisotropy energy) direction. This leads to an increase in the anisotropy energy. When such a growth is not compensated by the gain in energy due to the order parameter alignment with the fluctuations of the random field, the Imry-Ma inhomogeneous state becomes energetically unfavorable, and the system goes back to the state with the long range order.

The corresponding critical value was found in Ref. [3]

$$K_{cr} \sim J \left[\frac{x\langle h_l^2\rangle}{J^2 S^2}\right]^{\frac{2}{4-d}}. \tag{21}$$

It is easy to see that for *2<d<4* in the case of strongly anisotropic distribution of the random field directions, when the estimation (19) is valid, one has $K_{eff} \gg K_{cr}$, therefore the Imry-Ma inhomogeneous state is not implemented. The opposite condition should be fulfilled for its initiation.

On theoretical investigation of the system with weak uniaxial anisotropy, Aharony revealed [5] that increasing concentration of "random local field" type defects with the fields collinear to the easy axis leads to a phase transition to the phase where the order parameter is perpendicular to the easy axis. The author interpreted this transition as the defect-induced spin-flop transition. Taking into account above consideration, it would be more properly to refer it to the orientational transition. Indeed, the incorporation of defects reduces the magnitude of the anisotropy constant, and it changes its sign at critical concentration of defects, thus causing the occurrence of the "easy plane" type anisotropy.

## 5. Conclusions

It is established that in the case of anisotropic distribution of the directions of the defect-induced random local fields, the long-range order in the system with the original $O(n)$ symmetry and space dimensionality $2<d<4$ does not disappear due to the effective anisotropy induced by the defect fields in the order parameter space, and the Imry-Ma inhomogeneous state does not occur.